\documentclass[aps,superscriptaddress,amsmath,amssymb,twocolumn,amsfonts,floatfix, longbibliography]{revtex4-1}
\usepackage{mathtools}
\usepackage{braket}
\usepackage[dvipsnames]{xcolor}
\usepackage{float}
\usepackage{subfigure}
\usepackage{upgreek}
\usepackage{pgffor}
\usepackage{float}
\usepackage{silence}
\usepackage[colorlinks=true,linktoc=page,linkcolor=blue,citecolor=magenta,urlcolor=Fuchsia]{hyperref}
\usepackage[normalem]{ulem}
\usepackage{sidecap,tikz}
\usepackage{orcidlink}

\newcommand{\beq}{\begin{equation}}
\newcommand{\eeq}{\end{equation}}
\newcommand{\bea}{\begin{eqnarray}}
\newcommand{\eea}{\end{eqnarray}}

\newcommand{\eqn}[1] {Eq.~(\ref{#1})}
\newcommand{\fig}[1]{Fig.~\ref{#1}}
\newcommand{\sect}[1]{Sec.~\ref{#1}}
\newcommand{\mylabel}[1]{\label{#1}}

\mathchardef\mhyphen="2D 

\newcommand{\ie}{{\it i.e.,\,\,}}

\newcommand{\non}{\nonumber}  
\newcommand{\myfigwidth}{0.999\columnwidth}

\begin{document}
\title{Optimizing one dimensional superconducting diodes: Interplay of Rashba spin-orbit coupling and magnetic fields}	
	
\author{Sayak Bhowmik}\thanks{These authors contributed equally to this work}
\affiliation{Institute of Physics, Sachivalaya Marg, Bhubaneswar, Orissa 751005, India}
\affiliation{Homi Bhabha National Institute, Training School Complex, Anushakti Nagar, Mumbai 400094, India}

\author{Dibyendu Samanta}\thanks{These authors contributed equally to this work}
\affiliation{Department of Physics, Indian Institute of Technology, Kanpur 208016, India}

\author{Ashis K. Nandy}
\email{aknandy@niser.ac.in}
\affiliation{School of Physical Sciences, National Institute of Science Education Research,An OCC of Homi Bhabha National Institute, Jatni 752050, India}
	
\author{Arijit Saha}
\email{arijit@iopb.res.in}
\affiliation{Institute of Physics, Sachivalaya Marg, Bhubaneswar, Orissa 751005, India}
\affiliation{Homi Bhabha National Institute, Training School Complex, Anushakti Nagar, Mumbai 400094, India}

\author{{Sudeep Kumar Ghosh}\,\orcidlink{0000-0002-3646-0629}}
\email{skghosh@iitk.ac.in}
\affiliation{Department of Physics, Indian Institute of Technology, Kanpur 208016, India}
	
\date{\today}

\begin{abstract}
The superconducting diode effect (SDE) refers to the non-reciprocal nature of the critical current (maximum current that a superconductor can withstand before turning into a normal metal) of a superconducting device. Here, we investigate SDE in helical superconductors with broken inversion and time-reversal symmetry, focusing on a prototypical Rashba nanowire device proximitized by an $s$-wave superconductor and subjected to external magnetic fields. Using a self-consistent Bogoliubov-de Gennes mean-field formalism, we analyze the interplay between linear and higher-order spin-orbit coupling (SOC), bulk supercurrents, and external magnetic fields. Our results demonstrate that Rashba nanowires with only linear SOC can achieve incredibly large diode efficiency $\gtrsim 45\%$ through the interplay of longitudinal and transverse magnetic fields. Notably, higher-order SOC enables finite diode efficiency even without a longitudinal Zeeman field, which can be utilized to reveal its presence and strength in nanowires. We present a comprehensive phase diagram of the device elucidating the emergent Fulde-Ferrell-Larkin-Ovchinnikov (FFLO) superconducting state and demonstrate that proximitized Rashba nanowires offer a versatile, practical platform for SDE, with potential realizations in existing material systems. These results provide crucial insights for optimizing SDE in nanoscale superconducting devices, paving the way for next-generation dissipationless quantum electronics.
\end{abstract}
	
\maketitle

\section{Introduction}
\mylabel{sec:intro}
The invention of diodes marked a pivotal breakthrough in solid-state electronics, with their distinctive non-reciprocal current flow making them fundamental to semiconductor devices~\cite{Scaff_1947,Shockley_1949}. In quantum materials, inversion and time-reversal symmetry crucially determine charge current nonreciprocity. Magnetochiral anisotropy (MCA), arising in systems with broken inversion symmetry~\cite{Tokura2018}, is a dominant mechanism for this effect. MCA leads to non-reciprocity in the magnetoresistance with respect to the current direction and can be greatly enhanced by superconducting correlations~\cite{Qin2017,Wakatsuki2017}. In superconductors, MCA can lead to an extreme manifestation of nonreciprocity known as the superconducting diode effect (SDE), where resistance is zero in one current direction and non-zero in the other~\cite{Nadeem2023,Nagaosa2024,Narita2024}.  SDE thus realizes an ideal diode with perfect rectification. Such a scenario arises in the regime where the critical current that causes superconductor to normal metal transition differ in the opposite directions ($|J_{c}^+| \neq |J_{c}^-|$)~\cite{Nadeem2023,Nagaosa2024}. Recent experiments in engineered superlattice systems~\cite{Ando2020,sundaresh2023,Narita2024} and bulk materials/ thin films~\cite{Itahashi2020,Schumann2020,Wakatsuki2017,Nadeem2023} have spurred research interests into more efficient superconducting diode devices, including in twisted multilayer graphene~\cite{Lin2022,diez2023twisted_bi} and transition metal dichalcogenides~\cite{Bauriedl2022,Yun2023}.

The inversion symmetry breaking required for critical current non-reciprocity can be achieved through extrinsic mechanisms (e.g., asymmetric geometry or artificial superlattice potential~\cite{sundaresh2023,burlakov2014,Narita2024}) or intrinsic mechanisms like Rashba spin-orbit coupling (SOC)~\cite{Ando2020,Yuan_2022,ilic_2022,Turini2022,Roig2024,halterman_2022,akito_2022,karbassov_2022,Legg_2022}. Correspondingly, SDE can be broadly classified into two types: extrinsic and intrinsic. Although predicted decades ago~\cite{Levitov1985,Edelstein1996}, the mechanism of SDE is beginning to be understood only recently, leading to extensive investigations into its potential for dissipationless electronic circuits~\cite{Yuan_2022,Daido_2022,ilic_2022,He_2022,Roig2024,pal2022,Picoli_2023,Banerjee2024,Nadeem2023}. Recent theoretical studies~\cite{Yuan_2022,Daido_2022,ilic_2022,He_2022}, using phenomenological Ginzburg-Landau theory and mean-field analysis of the two-dimensional Rashba-Zeeman Hubbard model, have shown that intrinsic SDE can arise in systems with broken inversion and time-reversal symmetry. In the helical superconducting ground state, this occurs due to asymmetry in Cooper pair momentum in the Fulde-Ferrell-Larkin-Ovchinnikov (FFLO) state~\cite{Fulde1964,Larkin1965}, resulting in Kramers non-degenerate $s$-wave Cooper pairs~\cite{Daido_2022,Yuan_2022,Nadeem2023}.

A Rashba nanowire device proximitized to an s-wave superconductor and subjected to external magnetic fields serves as a prototypical platform for realizing helical superconductivity~\cite{Oreg2010,Sau2010,Lutchyn2010,Stanescu2011,Legg_2022, Picoli_2023}. While previous phenomenological/ non-self-consistent studies~\cite{Legg_2022} demonstrated the potential for SDE in such systems, they reported low diode efficiencies ($\sim 2\%$). 
To enhance the relatively low diode efficiency of this device and better understand its properties, it's crucial to systematically investigate the superconducting order parameter using self-consistent methods~\cite{Daido_2022,ilic_2022,Picoli_2023}. Recent studies~\cite{Daido_2022,ilic_2022,Picoli_2023} have shown that this approach can potentially uncover pathways for improvement across various parameters.
In this paper, we address this gap through a comprehensive, self-consistent mean-field investigation of the interplay between various Rashba SOC terms and external magnetic fields in determining emergent FFLO pairing and highly efficient superconducting diodes in a Rashba nanowire device. We show that even with only linear Rashba SOC we can achieve a very large diode efficiency ($\gtrsim 45\%$) by suitably tuning the parameters in this system. We extend beyond linear Rashba SOC to include higher-order (cubic) Rashba SOC terms, which are not only naturally present in semiconducting nanowires~\cite{Campos2018,Kammermeier2018,Sonehara2022} but also can dominate the normal state properties~\cite{PhysRevB.85.075404,PhysRevLett.113.086601,Nakamura2012,Michiardi2015,Shanavas2016,Liu2018,Peng2023,Amundsen2024}. Notably, we demonstrate that higher-order Rashba SOC can induce finite diode efficiency ($\gtrsim 15\%$) even without longitudinal magnetic fields. This behavior differs qualitatively from the linear SOC case and can be used to detect the presence of higher-order SOC terms in the nanowire. Our study establishes the Rashba nanowire device as a promising platform for highly efficient superconducting diodes, crucial for advancing nanoscale superconducting electronics.

The remainder of this article is structured as follows. \sect{sec:model} introduces the proximity-induced Rashba nanowire device and outlines the self-consistent mean-field formalism used to investigate properties of the superconducting ground state in the nanowire. \sect{sec:nonrec} presents our main findings on nonreciprocal charge transport in this device. We examine cases with both linear and higher-order Rashba SOC terms, exploring their interplay with Zeeman fields to optimize the SDE. Finally, \sect{sec:summ} summarizes our results, discusses potential material realizations, and proposes future research directions.


\begin{figure}[t]
\centering
\includegraphics[width=\myfigwidth]{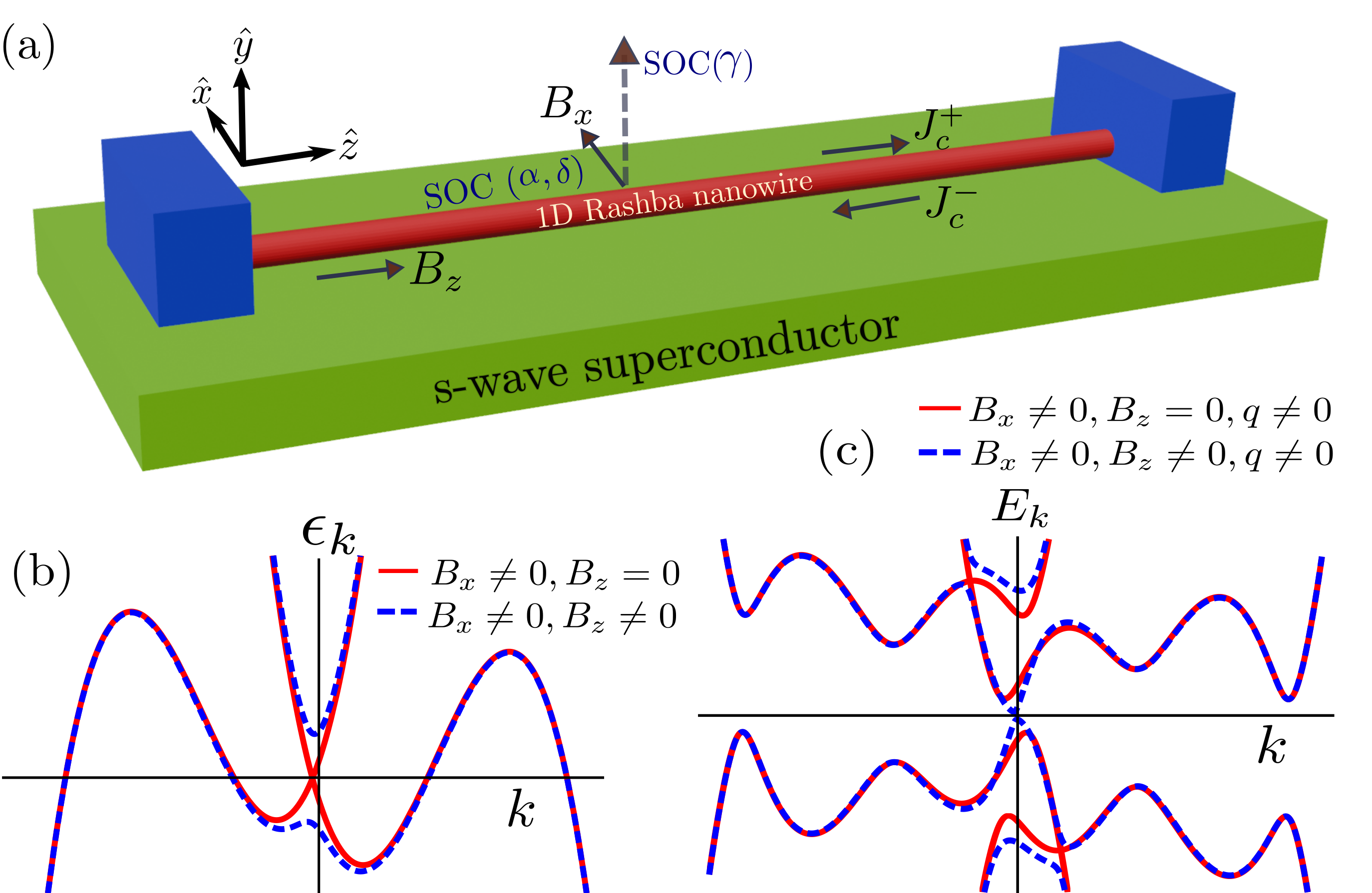}
\caption{\textbf{Schematic of the proximity-induced Rashba nanowire device:} (a) The schematic of our proposed device consisting of a Rashba nanowire in close proximity to an $s$-wave superconductor and in presence of external Zeeman fields: $B_x$ and $B_z$ to investigate SDE. The strength of the linear Rashba SOC is denoted by $\alpha$, while the cubic Rashba SOC strengths are represented by $\delta$ and 
$\gamma$ as illustrated in \eqn{eqn:nanowire_Ham}. (b) and (c): The dispersions of the normal state and the Bogoliubov quasi particles in the superconducting state are qualitatively shown in (b) and (c) respectively considering both linear and higher order SOC. 
}
\label{fig:schematic}
\end{figure}

\section{Model and Self-consistent mean-field formalism} 
\mylabel{sec:model}
We consider a single channel one-dimensional semiconducting nanowire along the $z$-direction with both linear and higher order Rashba SOC terms~\cite{Campos2018,Kammermeier2018,Sonehara2022,hsocnote} placed in close proximity to a three dimensional bulk $s$-wave superconductor as schematically shown in \fig{fig:schematic}(a). In the presence of externally applied magnetic fields, the normal state physics of the nanowire is governed by the one-dimensional Rashba-Zeeman Hamiltonian~\citep{Alicea2012}
\beq
\mathcal{H}_0 = \sum_{s,s^\prime} \int dz \, \psi_s^\dagger(z) h_{s s^\prime}(z)\psi_{s^\prime}(z),\non
\eeq
\vspace{-0.5cm}
\beq
\!\!\hat{h}(z) =  \frac{p_z^2}{2m} - \mu  + \left(\frac{\alpha}{\hbar} p_z + \frac{\delta}{\hbar^3} p_z^3\right)\sigma_x + \frac{\gamma}{\hbar^3} p_z^3 \sigma_y + \mathbf{B} \cdot \pmb{\sigma}\,.
\label{eqn:nanowire_Ham_realspace}
\eeq
Here, the spin degree of freedom is labeled by $s,s^\prime\in\{\uparrow,\downarrow\}$.  $\psi_s^\dagger(z)$ creates a fermion at the position $z$ with spin $s$, while $\pmb{\sigma} = (\sigma_x, \sigma_y, \sigma_z)$ is the vector of the Pauli matrices in spin space and $\mu$ is the chemical potential. $p_z = \frac{\hbar}{i} \partial_z$ is the momentum operator in the direction of current flow ($z$-direction) within the nanowire. The strength of the usual linear Rashba SOC is denoted by $\alpha$ and the cubic Rashba SOC of strengths are $\delta$ and $\gamma$. The presence of Rashba SOC breaks the inversion symmetry of the system. The externally applied magnetic fields lead to the Zeeman term characterized by $\mathbf{B} = (B_x, B_y, B_z)$ that breaks time-reversal symmetry. We assume the radius of the nanowire to be small such that there is only one occupied mode, \ie single channel. This nanowire Hamiltonian [\eqn{eqn:nanowire_Ham_realspace}] in momentum space can be written as
\beq
\mathcal{H}_0 = \sum_{s,s^\prime} \sum_k \, c_{k,s}^\dagger h_{s s^\prime}(k)c_{k,s^\prime},\non
\eeq
\vspace{-0.5cm}
\beq
\hat{h}(k) = \xi_k + \left[\left(\alpha k + \delta k^3\right)\sigma_x + \gamma k^3 \sigma_y\right] + \mathbf{B} \cdot \pmb{\sigma}\,. 
\label{eqn:nanowire_Ham}
\eeq
Here, $k$ represents the momentum along the nanowire ($z$-direction) and $\xi_k = \frac{\hbar^2 k^2}{2m}-\mu$. The bulk energy spectrum of the nanowire obtained by diagonalizing $\hat{h}_k$ 
is given by
\beq
\epsilon_\pm (k) = \xi_k \pm \sqrt{(\alpha k + \delta k^3 + B_x)^2 + (\gamma k^3 + B_y)^2 + B_z^2}\non
\eeq
where, $\pm$ label the two helicty bands. Note that, the Zeeman field $B_z$ (longitudinal field) opens a Zeeman gap while $B_x$ (transverse field) creates asymmetry between the two bands as schematically shown in 
Fig.~\ref{fig:schematic}(b). On the other hand, $B_y$ opens up a gap as well as creates band asymmetry (not shown). Throughout the majority of this paper, we will discuss the results with $B_y=0$ for simplicity and will separately discuss the effects of nonzero $B_y$ for completeness.

We seek to elucidate the nature of the superconducting ground state in the Rashba nanowire in the presence of $s$-wave superconducting correlations that can be induced via the proximity effect in presence of a bulk $s$-wave superconductor, as well as supercurrents (see \fig{fig:schematic}(a)). In general, the presence of supercurrents can lead to a superconducting state with finite momentum Cooper pairs termed 
as the FFLO state. To establish that, 
we consider an attractive Hubbard type interaction in the bulk $s$-wave superconductor described by the Hamiltonian
\beq
\mathcal{H}_I=-\frac{U}{2}\sum_{s,s^\prime} \int d^{3}{\bf{r}} \, \psi_{s}^\dagger({\bf{r}})\psi_{s^\prime}^\dagger({\bf{r}})\psi_{s^\prime}({\bf{r}})\psi_{s}({\bf{r}})\,,
\eeq
where, $U$ is the attractive Hubbard interaction strength. 
We now decouple the Hubbard interaction term ($\mathcal{H}_I$) in the finite momentum $s$-wave channel within the mean-field approximation~\cite{Daido_2022,akito_2022,Picoli_2023} and consider these correlations to be proximity induced in the nanowire. Then, using the Bogoliubov de-Gennes (BdG) mean-field formalism the effective Hamiltonian of the nanowire takes the form     
\bea
\mathcal{H} & = & \frac{1}{2}\int\,dz\,\Psi^\dagger(z)\mathcal{H}_{BdG}(z)\Psi(z) + \mathcal{E}_0 \,,\non \\
\!\!\!\!\!\!\!\!\mathcal{H}_{BdG} (z) & = & 
\begin{bmatrix}
\hat{h}(z) & \hat{\Delta}(z) \\
-\hat{\Delta}^*(z) & -\hat{h}^*(z)
\end{bmatrix}; \, \hat{\Delta}(z)  =   -i\sigma_y \Delta  e^{iqz}\,,
\mylabel{eqn:BdG_Ham}
\eea
where, $\Psi(z)=[\psi_{\uparrow}(z),\psi_{\downarrow}(z),\psi_{\uparrow}^\dagger(z),\psi_{\downarrow}^\dagger(z)]^T$ is the Nambu spinor inside the nanowire, $\mathcal{E}_0 = \frac{L}{U}|\Delta|^2$ 
with $L$ denoting the length of the nanowire. The $s$-wave FFLO order parameter is given by $\Delta(z)= \Delta (-i\sigma_y) e^{iqz}$, where $q$ 
is the Cooper pair momentum. The Hamiltonian in \eqn{eqn:BdG_Ham} effectively describes the physics of the Rashba nanowire device schematically shown in \fig{fig:schematic}(a). To proceed further, we write the mean-field Hamiltonian [\eqn{eqn:BdG_Ham}] in the momentum space as
\bea
\mathcal{H} &=& \frac{1}{2} \sum_{k}\Psi_{k}^\dagger\mathcal{H}_{BdG} (k) \Psi_{k}+ \mathcal{E}_0 \,, \non\\
&=& \sum_{n,k} \left(E_{n,k} \gamma^\dagger_{n,k} \gamma_{n,k} - \frac{E_{n,k}}{2} \right) + \mathcal{E}_0 \,, \non 
\eea
\vspace{-0.3cm}
\beq
\text{\rm with,}\;\;\mathcal{H}_{BdG} (k) =
\begin{bmatrix}
\hat{h}(k + \frac{q}{2}) & -i\sigma_y \Delta \\
i\sigma_y \Delta & - \hat{h}(-k+\frac{q}{2})^*
\end{bmatrix}\,.
\eeq
Here, $\Psi_{\mathbf{k}} \equiv [c_{k+q/2,\uparrow},c_{k+q/2,\downarrow},c_{-k+q/2,\uparrow}^\dagger,c_{-k+q/2,\downarrow}^\dagger]^T$ and $E_{n,k} (q)$ denotes the Bololiubov quasiparticle energies. 
The behavior of the Bogoliubov quasiparticle spectrum is schematically shown in \fig{fig:schematic}(c). This depicts that the introduction of the pairing potential $\Delta$ opens a gap in the spectrum. 
As in the case of the nanowire spectrum, the asymmetry in BdG bands occurs due to the application of the Zeeman field $B_x$, 
and the Zeeman field $B_z$ makes the bands distinctly gapped.

\begin{figure}[!htb]
\centering 
\includegraphics[width=\linewidth]{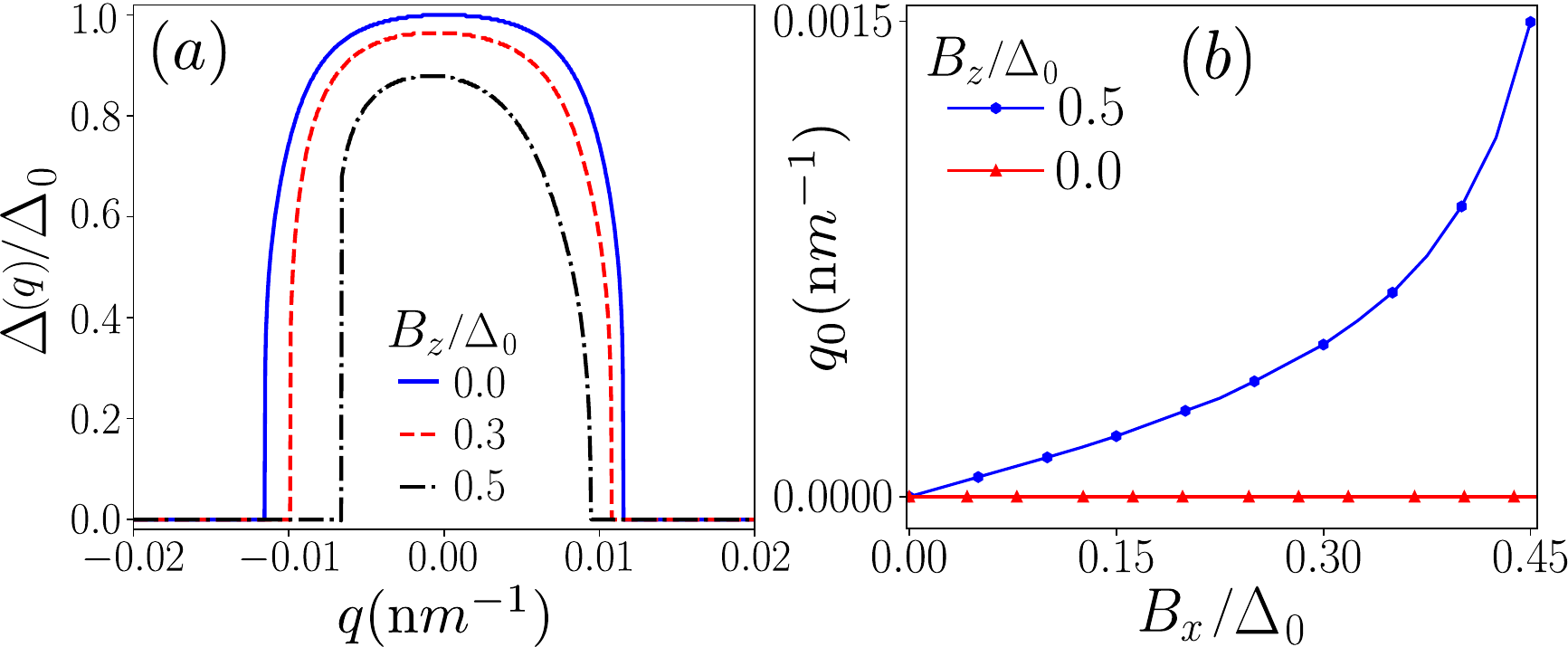}
\caption{\textbf{FFLO ground state with linear SOC:} (a) The self-consistent superconducting gap $\Delta (q)$ is shown as a function of the Cooper-pair momentum $q~$(nm\textsuperscript{-1}) for $B_x/\Delta_0=0.3$ with only linear Rashba SOC ($\alpha = 100$ meV-nm). $\Delta_0$ is the BCS gap in the absence of any external magnetic field ($\mathbf{B}=0$). (b) The Cooper pair momentum of the FFLO ground state $q_0$ as a function of $B_x$. The presence of the FFLO ground state ($q_0 \neq 0$) is established only when both $B_x$ and $B_z$ are finite.  Other model parameters are chosen as: $(B_y,\mu/\Delta_0, U, \beta^{-1})$=($0$, $0.5$, $25.55$ \text{meV}, $0.1$ \text{meV}).}   
\label{fig:lsoc_fflo}
\end{figure}

\begin{figure}[!b]
\centering 
\includegraphics[width=\linewidth]{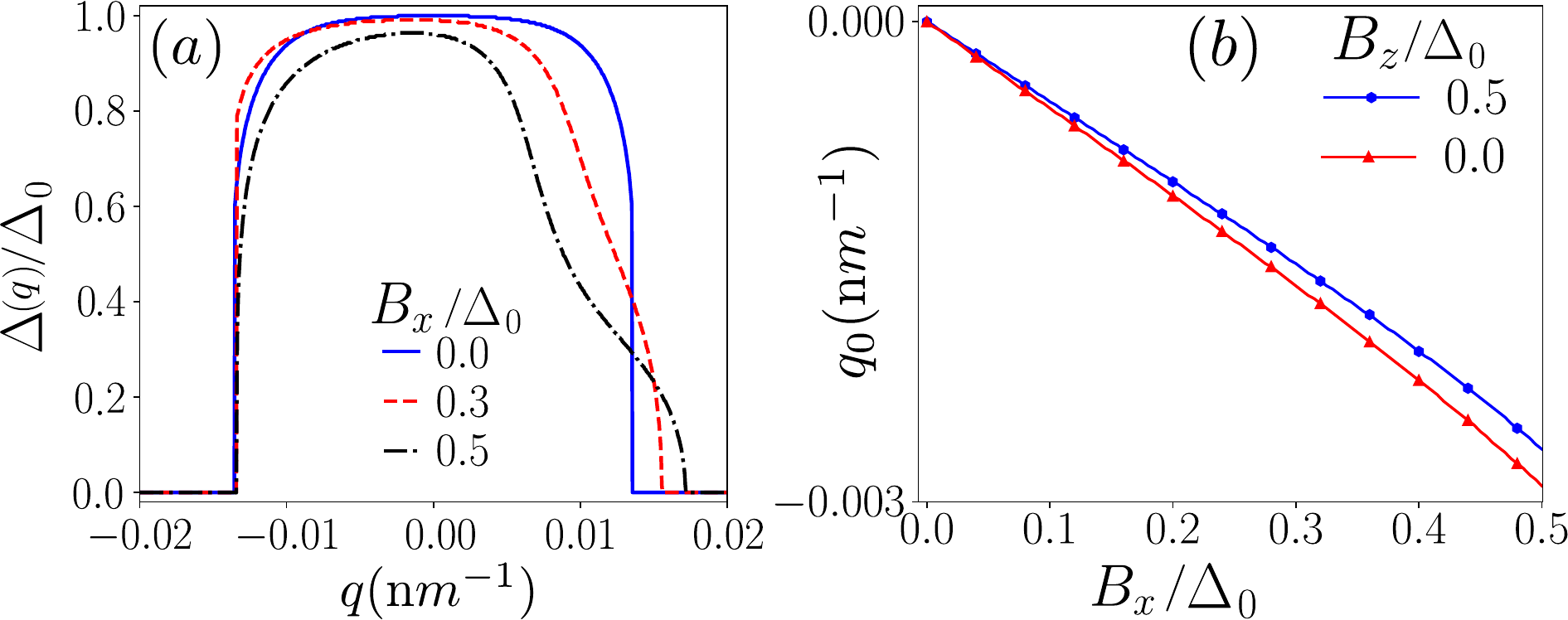}
\caption{\textbf{FFLO ground state with both linear and higher order SOC:} (a) The self-consistent superconducting gap $\Delta (q)$ is depicted as a function of Cooper-pair momentum $q~$ for $B_z=0$ with both the linear Rashba SOC term ($\alpha=100$ meV-nm) and the cubic Rashba SOC terms ($\delta = \gamma = 10$ eV-nm\textsuperscript{3}). (b) The  Cooper pair momentum of the FFLO ground state $q_0~$ as a function of $B_x$ is shown. Note that the FFLO state is stable even when $B_{z}=0$ in the presence of a finite $B_x$. Other system parameters are chosen as: $(B_y, \mu/\Delta_0, U, \beta^{-1})$=($0$, $0.5$, $16.45$ 
\text{meV}, $0.1$ \text{meV}).}   
\label{fig:csoc_fflo}
\end{figure}

The condensation energy $\Omega(q)$ of the system, defined as the difference of free energy per unit length between the superconducting and normal state~\cite{Kinnunen_2018, Daido_2022}, is
\beq
\Omega(q,\Delta) = F(q,\Delta) - F(q,0)\,,
\label{eqn:cond_energy}
\eeq
where, $F(q,\Delta)$ is the free energy density of the nanowire given by
\beq
F(q,\Delta) = -\frac{1}{L\beta}\sum_{n,k} \ln\left[1+e^{-\beta E_{n,k}(q)}\right] + \frac{|\Delta(q)|^2}{U}\,,
\label{eqn:free_energy}
\eeq
where, $\beta=(k_B T)^{-1}$ with $k_B$ being the Boltzmann constant and $T$ is the temperature.

\begin{figure*}[!htb]
\centering 
\includegraphics[width=\linewidth]{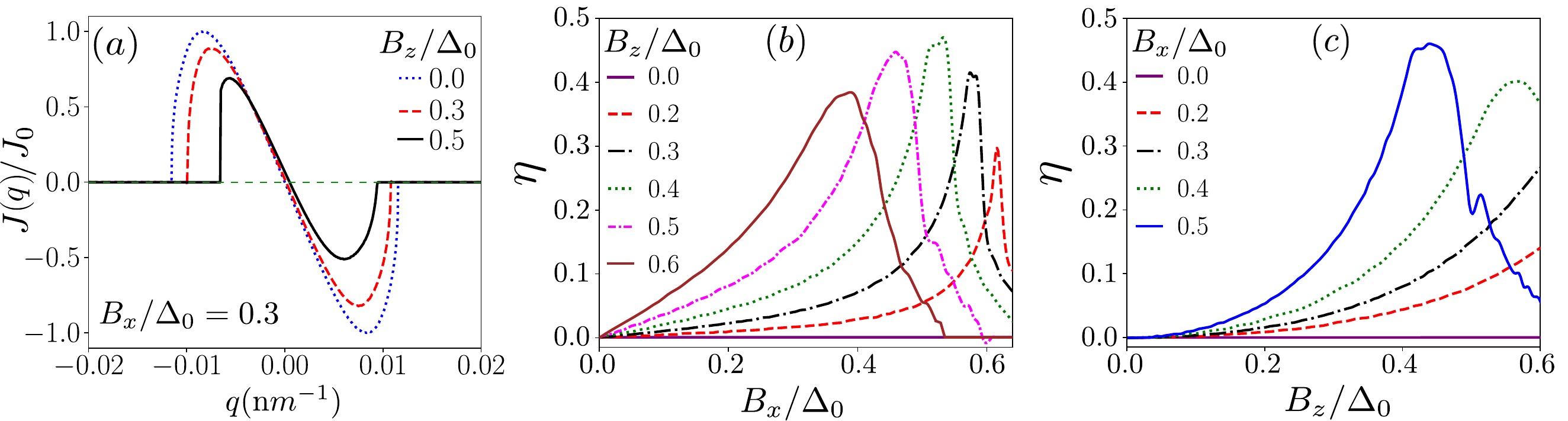}
\caption{\textbf{Superconducting diode effect with linear SOC:} (a) The supercurrent density ($J$) is shown as a function of $q~$(nm\textsuperscript{-1}) considering solely linear Rashba SOC and $B_x/\Delta_0=0.3$. The nonreciprocity in the supercurrent, $J(q)\neq -J(-q)$, is achieved in this case only when both $B_x$ and $B_z$ are finite. It then leads to nonreciprocity in the critical supercurrents, $J_{c}^+ \ne|J_{c}^-|$, hence exhibiting a finite diode efficiency ($\eta \neq 0$). Here, $J_0 \equiv J_c (B_z = 0) = |J_{c}^+ (B_z = 0)| = |J_{c}^- (B_z = 0)|$. (b) and (c): The corresponding diode efficiency ($\eta$) is shown as a function of $B_x$ (with fixed $B_z$) and $B_z$ (with fixed $B_x$) with only linear Rashba SOC in the panels (b) and (c) respectively. The system parameters are chosen as: $(B_y, \alpha, \mu/\Delta_0, U, \beta^{-1})$=($0$, $100$ meV-nm, $0.5$, $25.55$ \text{meV}, $0.1$ \text{meV}).}   
\label{fig:lsoc_efficiency}
\end{figure*}

We determine the order parameter $\Delta(q)$ for a given value of $q$ by self-consistently solving the gap equation, obtained via minimizing the free energy $F(q,\Delta)$ in \eqn{eqn:free_energy}, given by
\beq
\Delta(q) = -\frac{U}{L} \sum_{n,k} \frac{\partial E_{n,k}}{\partial \Delta^*} \,n_F(E_{n,k})
\eeq
where $n_F(E_{n,k}) = 1/(1 + e^{\beta E_{n,k}})$ is the Fermi function. Using the self-consistent solution of $\Delta(q)$ in \eqn{eqn:cond_energy}, we optimize the condensation energy with respect to $q$ to obtain the Cooper pair momentum ($q_0$) of the true FFLO superconducting ground state. The BCS gap in the absence of any external magnetic fields ($\mathbf{B}=0$) is denoted by $\Delta_0$ and we choose the value of $U$ in such a way that we are in the weak coupling BCS regime with $\Delta_0 \sim 1$ meV. In the rest of the paper, we present results in the low temperature limit choosing a fixed value of $\beta^{-1}=0.1$ meV much smaller than $\Delta_0$.

The variation of the self-consistent solutions of the superconducting order parameter $\Delta(q)$ and the momentum of the Cooper pairs $q_0$ in the FFLO ground state with the external Zeeman fields are shown in \fig{fig:lsoc_fflo} and \fig{fig:csoc_fflo} with only linear SOC case and, with both the linear and higher order SOC case respectively. We note from Fig~\ref{fig:lsoc_fflo}(a) and Fig~\ref{fig:csoc_fflo}(a) that for both the cases the superconducting gap vanishes for a critical value of the Cooper pair momentum either above a positive value of $q = q_{c}^+$ or below a negative value of $q = q_{c}^-$. Fig~\ref{fig:lsoc_fflo}(b) and Fig.~\ref{fig:csoc_fflo}(b) show that $q_0$ varies linearly with $B_x$ for $B_x \ll \Delta_0 $ for a given value of $B_z$. We note Fig~\ref{fig:lsoc_fflo}(b) that in the case of only linear SOC $q_0$ is finite only when $B_z \neq 0$ and $B_x \neq 0$. This is because when only linear SOC is present the effect of the SOC can be gauged away by using a spin-dependent gauge transformation~\cite{Braunecker2010} and then when only $B_x \ne 0$ the band structure is symmetric leading to zero momentum Cooper pairs. However, when both $B_x$, $B_z \ne 0$, then in the transformed basis a spiral magnetic field texture is realized leading to momentum dependent asymmetry in the band structure and hence leading to finite momentum Cooper pairs. In contrast, when both the linear and cubic SOC terms are present, the effect of SOC can not be  gauged away and as a result only $B_x \neq 0$ is sufficient to give rise to $q_0 \neq 0$ in this case. Consequently, this feature also gets manifested in realizing finite SDE in this Rashba nanowire device, as extensively discussed in the next section.

\begin{figure*}[!htb]
\centering 
\includegraphics[width=\linewidth]{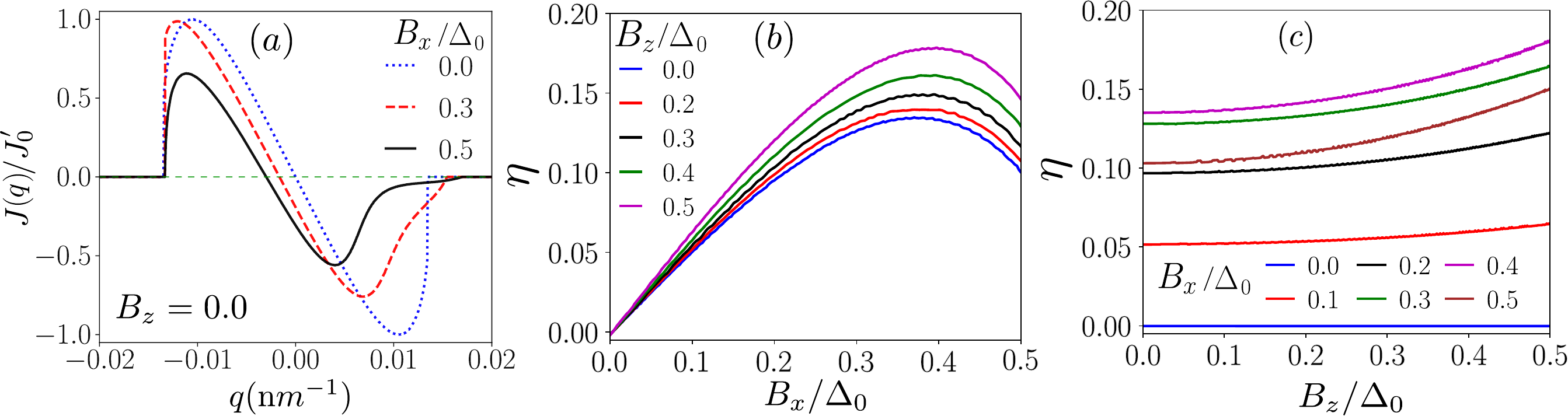}
\caption{\textbf{Superconducting diode effect with both linear and higher order SOC:} (a) The supercurrent density ($J$) as a function of $q~$(nm\textsuperscript{-1}) with both cubic Rashba SOC ($\delta$ = $\gamma$ = $10$ eV-nm\textsuperscript{3}) and linear Rashba SOC ($\alpha=100$ meV-nm) choosing $B_z=0$. The nonreciprocity in the critical supercurrents: $|J_{c}^+|\ne|J_{c}^-|$ (also implying a finite diode efficiency $\eta \neq 0$) is achieved for any finite value of $B_x$. Here, $J'_0 \equiv J_c (B_x = B_z = 0) = |J_{c}^+ (B_x = B_z = 0)| = |J_{c}^- (B_x = B_z = 0)|$. (b) and (c): The corresponding diode efficiency ($\eta$) is depicted as a function of $B_x$ (with fixed $B_z$) and  $B_z$ (with fixed $B_x$) in the panels (b) and (c) respectively. We have chosen the other model parameters as: $(B_y,\mu/\Delta_0, U, \beta^{-1})$=($0$, $0.5$, $16.45$ \text{meV}, $0.1$ \text{meV}).}   
\label{fig:csoc_efficiency}
\end{figure*}

\section{Non-reciprocal transport}
\mylabel{sec:nonrec}
The supercurrent density $J(q)$ is computed from the condensation energy $\Omega(q,\Delta)$ in \eqn{eqn:cond_energy} as~\cite{Yuan_2022,Daido_2022,Legg_2022}, 
\begin{equation}
J(q)=-2e\,\frac{\partial\Omega(q,\Delta)}{\partial q}.
\label{SDEcurrent}
\end{equation}
The sign of $J(q)$ indicates the direction of the supercurrent flow. The critical (optimal) values of the supercurrent along and opposite to the flow direction within the range $q_{c}^- \leq q \leq q_{c}^+$ for which the superconducting phase is stable, are denoted as $J_{c}^+$ and $J_{c}^-$ respectively. When $|J_{c}^+| \neq |J_{c}^-|$, we have a finite SDE in the system and the efficiency or the quality factor of the SDE is defined as
\begin{equation}
\eta=\frac{|J_{c}^+|-|J_{c}^-|}{|J_{c}^+|+|J_{c}^-|}\ .
\end{equation}
We first compute $J(q)$ over the range $q_{c}^- \leq q \leq q_{c}^+$ using the self-consistent solutions $\Delta(q)$ for each value of $q$ (shown in Fig~\ref{fig:lsoc_fflo}(a) and Fig~\ref{fig:csoc_fflo}(a) for example). Then finding its optimal values $J_{c}^+$ and $J_{c}^-$, we evaluate the diode efficiency ($\eta$) numerically. We systematically analyze the SDE in our Rashba nanowire device considering two broadly distinct scenarios: with only linear SOC and, with both linear and higher order SOC for various system parameters as discussed in the following two subsections. 


\subsection{Case-I: Rashba nanowire with only linear SOC}
In the presence of only linear Rashba SOC, we uncover the SDE by investigating the behavior of the supercurrent density ($J(q)$) and the diode efficiency ($\eta$) for different system parameters as presented in \fig{fig:lsoc_efficiency}. \fig{fig:lsoc_efficiency}(a) shows that the supercurrent density is symmetric with respect to $q$ for $B_z = 0$ even with $B_x \neq 0$~\cite{Legg_2022}. We note that as we start increasing the magnitude of $B_z$ with $B_x \neq 0$, $J(q)$ becomes more and more asymmetric, leading to nonreciprocal critical supercurrents, $|J_{c}^+|\neq |J_{c}^-|$ and hence to finite SDE. This is a direct consequence of the fact that the presence of FFLO state and the asymmetry in critical momenta: $|q_{c}^+|\neq |q_{c}^-|$ is distinctly realized only in presence of both the fields $B_x, B_z$ as noted in \fig{fig:lsoc_fflo}(a). 

The diode efficiency as a function of $B_x$ for fixed values of $B_z$ and as a function of $B_z$ for fixed values of $B_x$ are shown in \fig{fig:lsoc_efficiency}(b) and \fig{fig:lsoc_efficiency}(c) respectively. In both cases, $\eta$ increases linearly at low field strengths, reaches a maximum at moderate values, and then decreases to zero as the field strengths continue to intensify. This behavior can be understood as a consequence of a competition between the asymmetry in the band structure of the system and changes in the superconducting pairing susceptibility as the magnetic field strengths are varied, consistent with previous studies~\cite{Yuan_2022,Turini2022}. Note that, the self-consistent BCS gap vanishes for very large values of the Zeeman fields $B_x$ and $B_z$ consequently leading to zero efficiency, however the resultant nanowire spectrum might still be gapped like an ordinary insulator. 

	
Preceding investigations on a Rashba nanowire featuring only linear Rashba SOC~\cite{Legg_2022} reported maximum diode efficiency of $\sim 2\%$, where the superconducting order parameter has been fixed to a particular value. Contrastingly, our study reveals that the self-consistently derived superconducting order parameter within the full parameters space has significant influence on the performance of the diode where the highest $\eta$ value is found to be $\gtrsim 45\%$. This is evident from calculation presented in \fig{fig:lsoc_efficiency}(b,c). The vanishing of the superconducting gap within our self-consistent treatment necessitates restricting the maximum values of the magnetic fields used in our model. We note that in this case $B_y$ also plays a similar role as $B_z$ due to the geometry of the setup (see \fig{fig:schematic}(a)) under consideration and hence further analysis considering $B_y \neq 0$ is not presented. However, $B_y$ plays a significant and distinct role when we consider the effect of cubic Rashba SOC terms as discussed in the next subsection.

\subsection{Case-II: Rashba nanowire with both linear and higher order SOC}
The variation of $J(q)$ and $\eta$ in the case when both the linear ($\alpha \ne 0$) and cubic Rashba SOC terms ($\delta,\gamma\neq 0$) are present in the nanowire are shown in \fig{fig:csoc_efficiency}. First of all, comparing \fig{fig:csoc_efficiency}(a) with \fig{fig:lsoc_efficiency}(a), we note that the cubic Rashba SOC terms interestingly lead to asymmetry in $J(q)$ even with $B_z = 0$ as long as $B_x \neq 0$. This leads to nonreciprocal critical supercurrents, $|J_{c}^+|\neq |J_{c}^-|$ with a single magnetic field $B_x \neq 0$ in contrast to only linear SOC case discussed earlier. The variation of $\eta$ with the magnetic fields $B_x$ and $B_z$ with $B_y =  0$ are shown in \fig{fig:csoc_efficiency}(b, c). We note from \fig{fig:csoc_efficiency}(b) that the diode efficiency at a fixed $B_z$ exhibits a linear increase as a function of $B_x$ for small values of $B_x \ll \Delta_0$, eventually attaining a maxima around $B_x/\Delta_0=0.4$ similar to the case with only linear Rashba SOC described earlier. This behavior, as in the earlier linear SOC case~\cite{Yuan_2022,Turini2022}, results from two competing effects: the system's asymmetric band structure versus changes in superconducting pairing susceptibility as magnetic field strength varies. The maximum efficiency increases as one increases the magnitude of $B_{z}$. However, $\eta$ shows only marginal enhancement at a fixed value of $B_x$ when $B_z$ is varied as depicted in \fig{fig:csoc_efficiency}(c). 

\begin{figure}[!t]
\centering 
\includegraphics[width=\linewidth]{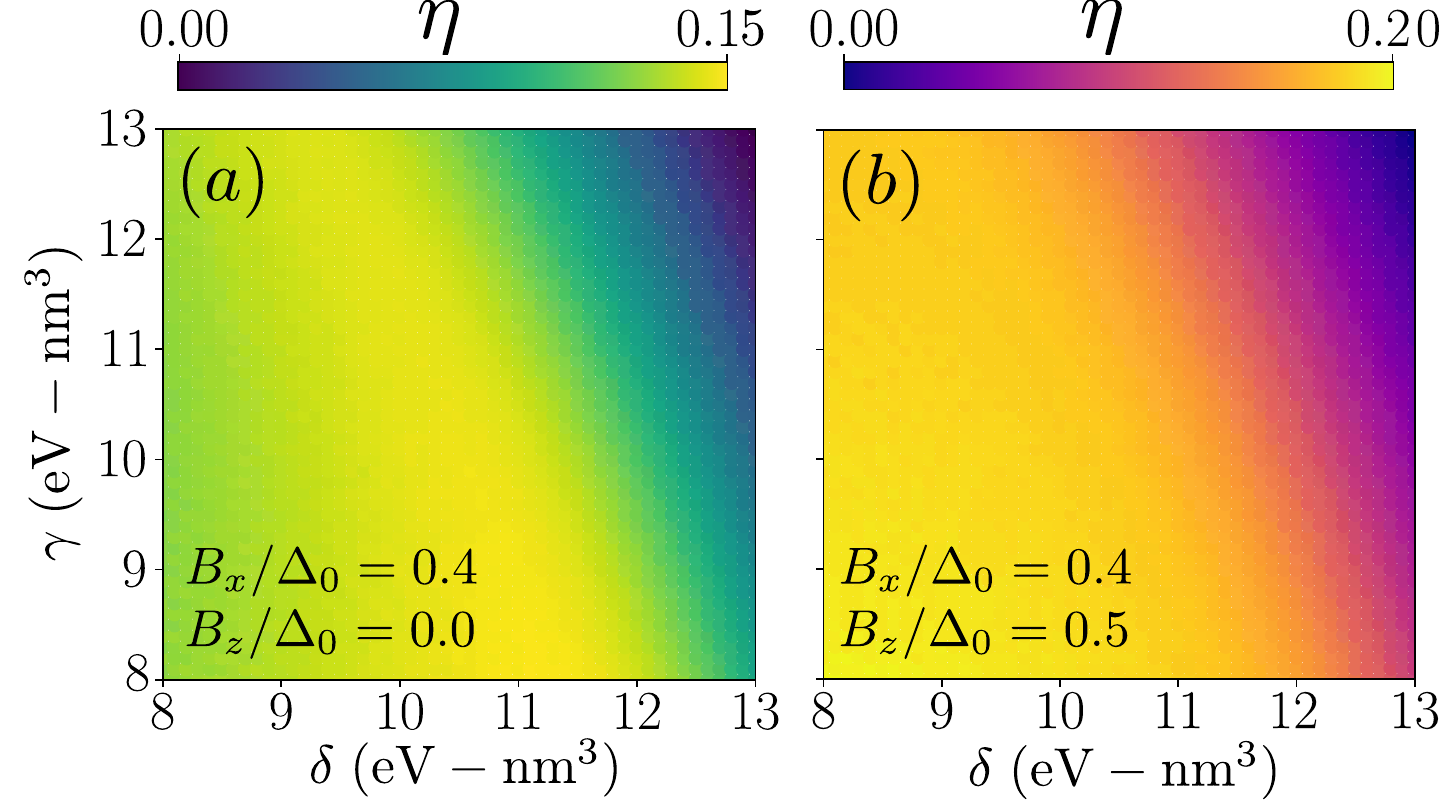}
\caption{\textbf{Diode efficiency phase diagram with both linear and higher order SOC:} The variation of the diode efficiency $\eta$ is shown in the plane of cubic Rashba SOC parameters $\delta - \gamma$ choosing ``optimal'' values of the Zeeman fields  $(B_x/\Delta_0, B_z/\Delta_0)=(0.4, 0.0)$ in panel (a) and 
$(B_x/\Delta_0, B_z/\Delta_0)=(0.4, 0.5)$ in panel (b), respectively. 
We note a small enhancement in the maximum value of $\eta$ for finite $B_z$. The  other system parameters are chosen as: $(B_y,\alpha, \mu/\Delta_0, \beta^{-1})$=$(0$, $100$ meV-nm, $0.5$, $0.1$ meV).}   
\label{fig:zeeman_csoc_efficiency}
\end{figure}

\begin{figure}[!b]
\centering 
\includegraphics[width=\linewidth]{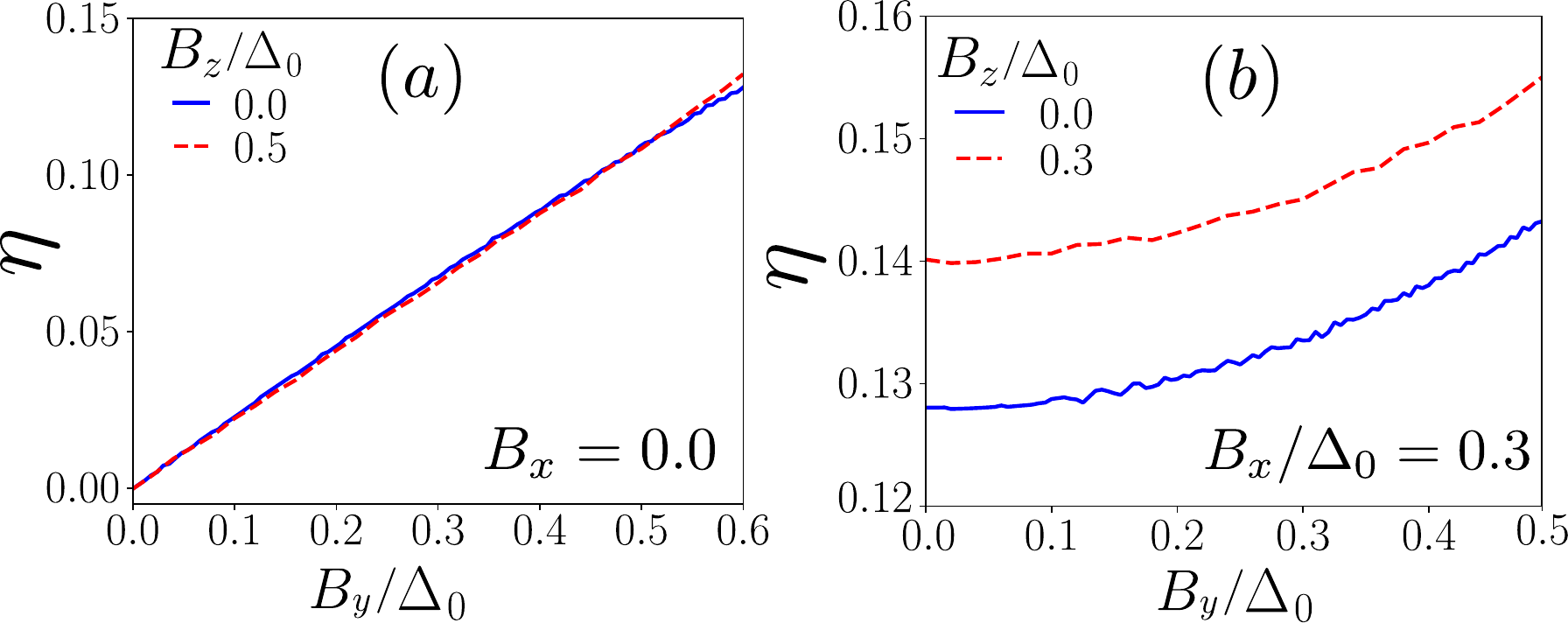}
\caption{The diode efficiency $\eta$ as a function of $B_y$ 
with both linear and cubic Rashba SOC terms for $B_x = 0$ and finite $B_x/\Delta_0 = 0.3$ are shown in (a) and (b) respectively. The chosen model parameters are $(\alpha, \delta,\gamma, \mu/\Delta_0, \beta^{-1})$=($100$ meV-nm, $10$~eV-nm\textsuperscript{3}, $10$~eV-nm\textsuperscript{3}, $0.5$, $0.1$ \text{meV}).}     
\label{fig:all_efficiency}
\end{figure}

Modifying the values of $\delta$ and $\gamma$ evidently changes the band dispersions in the normal and superconducting state of the system and consequently the efficiency of the superconducting diode will also change. We depict the diode efficiency phase diagrams in the $\delta-\gamma$ plane in Fig.~\ref{fig:zeeman_csoc_efficiency}. We obtain the phase diagrams by varying $U$ for every configuration of ($\delta,\gamma$) such that the self-consistent BCS gap $\Delta_0$ remains fixed. For the $B_z=0$ and $B_x \neq 0$ case, we can achieve a peak diode efficiency of approximately $15\%$ for an optimal choice of higher-order Rashba SOC parameters as shown in \fig{fig:zeeman_csoc_efficiency}(a). We emphasize that this behavior is qualitatively different from the linear SOC case, where the efficiency is zero with only the $B_x$ field. This distinctive response, arising solely from higher-order SOC terms, provides a method to detect and quantify the strength of these terms in a nanowire device.

The application of an additional magnetic field $B_z$ readily elevates $\eta$ to approximately $20\%$, as illustrated in Fig.~\ref{fig:zeeman_csoc_efficiency}(b) since effects of both the linear and cubic SOC terms are relevant. Note that the presence of highly asymmetric behavior in the supercurrent, given a specific set of system parameters, does not necessarily guarantee a corresponding high level of non-reciprocity in the super-currents. As is evident from \fig{fig:zeeman_csoc_efficiency}(a,b), the system might exhibit nearly negligible diode effect ($\eta \sim 0$) even though the parameters are set such that there is asymmetry in supercurrent (top right corners of \fig{fig:zeeman_csoc_efficiency}(a,b)).

Note that the results discussed so far were obtained by fixing $B_y = 0$. The effect of finite $B_y$ in $\eta$ is shown in the \fig{fig:all_efficiency}. In this case, $B_y$ serves the role of both $B_x$ and $B_z$ in realizing a finite SDE by creating both gap and asymmetry in the band structure. \fig{fig:all_efficiency}(a) and \fig{fig:all_efficiency}(b) show that the application of $B_y$ produces systematic increase in $\eta$ with increasing $B_y$ for different values of $B_z$ with $B_x = 0$ and $B_x \neq 0$ respectively. However, $B_y$ can not increase the efficiency beyond a certain limit because the superconducting state becomes unstable after some value of $B_y$ when the other magnetic fields are present as well.

\begin{figure}[!t]
\centering 
\includegraphics[width=\linewidth]{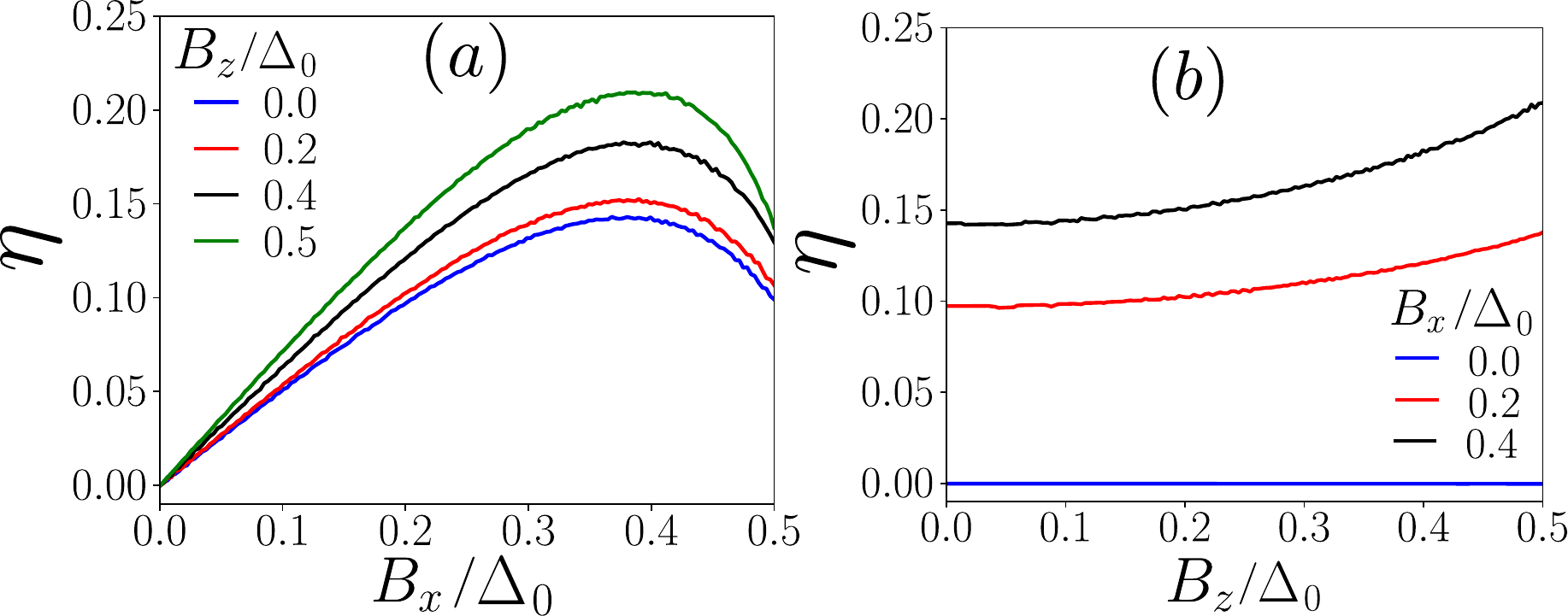}
\caption{The diode efficiency ($\eta$) with linear SOC and with cubic SOC of different type where $\delta= 10$~eV-nm\textsuperscript{3} but $\gamma=0$ as a function of $B_x$ (with fixed $B_z$) and  $B_z$ (with fixed $B_x$) are shown in (a) and (b) respectively. Other parameters are: $(\alpha, \mu/\Delta_0, B_y, \beta^{-1})$=($100$ meV-nm, $0.5$, $0$, $0.1$ \text{meV}).}     
\label{fig:c2_symm_efficiency}
\end{figure}

For completeness, we consider a special case with a different type of the higher order Rashba SOC terms~\cite{hsocnote} when only $\delta \ne 0$ and $\gamma=0$ (with $\alpha \ne 0$ as usual). The variation of the corresponding diode efficiencies with the magnetic fields are shown in \fig{fig:c2_symm_efficiency}. A finite diode efficiency with a maximum of $\sim 12\%$ even in presence of a finite $B_x$ only is obtained as shown in \fig{fig:c2_symm_efficiency}(a) which is also in contrast to the linear SOC case. Thus the presence of even a single cubic Rashba parameter $\delta$ is sufficient to manifest SDE. By further application of $B_z$ the diode efficiency $\eta$ is readily enhanced upto $\sim 21\%$ as a function of $B_x$ shown in \fig{fig:c2_symm_efficiency}(a). There is only a marginal increase in efficiency by increasing $B_z$ for fixed $B_x$ as shown in the \fig{fig:c2_symm_efficiency}(b).

\section{Summary and conclusions} 
\mylabel{sec:summ}
This article analyzes how the interplay between linear and cubic Rashba SOC terms, combined with external Zeeman fields, can optimize the efficiency of SDE in a proximitized one-dimensional nanowire device. Using self-consistent mean-field analysis, we demonstrate that SDE, characterized by nonreciprocal critical currents ($|J_{c}^+| \neq |J_{c}^-|$), arises from Cooper pair momentum asymmetry in the helical superconducting ground state of this time-reversal and inversion symmetry broken system. Remarkably, we achieve a diode efficiency $\eta \gtrsim 45\%$ with only linear Rashba SOC significantly higher than the previously reported efficiency of $\eta \sim 2\%$ from non-self-consistent analyses~\cite{Legg_2022}. In this case with only linear SOC, at least two magnetic fields are required to obtain a finite diode efficiency: one along the SOC direction and one perpendicular to it. This requirement stems from the fact that linear SOC effects can be eliminated through a spin-dependent gauge transformation~\cite{Braunecker2010}, resulting in a conventional BCS superconducting state with zero Cooper pair momentum when only one Zeeman field is present. However, the presence of two orthogonal Zeeman fields creates a spiral magnetic field texture via the gauge transformation, leading to momentum-dependent band asymmetry and asymmetric FFLO Cooper pairs~\cite{Braunecker2010}. In contrast, higher-order SOC effects cannot be eliminated via gauge transformations, allowing for finite diode efficiency even with only one Zeeman field unlike the linear SOC case and this behavior can be used to detect the presence of higher-order SOC in nanowires. However, including higher-order SOC doesn't improve the maximum diode efficiency (only $\eta \sim 20\%$), this value remains significant compared to previously reported diode efficiencies~\cite{Legg_2022,Nadeem2023}. Our results underscore the importance of self-consistently determining the order parameter of the superconducting ground state to fully understand the capability and optimize Rashba nanowire devices as superconducting diodes.

Our findings can be readily experimentally tested in spin orbit coupled Rashba nanowires~\citep{Alicea2012,Liang2012,Manchon2015,Bercioux2015} such as zinc-blende InSb and wurtzite InAs, with lengths of approximately 100 nm, where cubic Rashba SOC interactions are significant depending on the growth direction~\cite{Campos2018}. Superconductivity can be proximity induced in these nanowires by common bulk $s$-wave superconductors like Nb or Al. Future research directions include investigating SDE in proximity-induced multichannel Rashba nanowires with overlapping channels~\cite{Potter2010,Lutchyn2011,Law2011,Alicea2012}, extending beyond the single-channel case examined here. Additionally, the influence of topological superconductivity realized in the nanowire and the nanowire device geometry on SDE warrants further exploration~\cite{Liu2023,Kutlin2020,Kopasov2021,Legg2023P,Nesterov2016,Spansl2018}. An intriguing prospect is the potential replication of magnetic field effects using electric fields via the Rashba-Edelstein effect~\cite{Vignale2016,Smirnov2017,Piatti2021,Nadeem2023}, which could lead to the realization of field-free SDE.

\section{Acknowledgments} 
SKG acknowledges financial support from SERB, Government of India via the Startup Research Grant: SRG/2023/000934 and IIT Kanpur via the Initiation Grant (IITK/PHY/2022116). A.K.N. acknowledges the financial support from Department of Atomic Energy (DAE), Govt. of India, through the project Basic Research in Physical and Multidisciplinary Sciences via RIN4001. The authors acknowledge discussions with Koushik Mandal, Krishnendu Sengupta and Amit Agarwal. DS and SKG utilized the \textit{Andromeda} server at IIT Kanpur for numerical calculations. SB and AS acknowledge SAMKHYA: High-Performance Computing Facility provided by Institute of Physics, Bhubaneswar, for numerical computations.

\bibliography{sde}
\end{document}